# Inverse Z-spectrum analysis for MT- and spillover-corrected and $T_1$-compensated steady-state pulsed CEST-MRI – application to pH-weighted MRI of acute stroke

**Moritz Zaiss[1], Junzhong Xu[2,3], Steffen Goerke[1], Imad S. Khan[4], Robert J. Singer[4], John C. Gore[2,3,5], Daniel F. Gochberg[2,3,6], Peter Bachert[1]**

[1]Department of Medical Physics in Radiology, Deutsches Krebsforschungszentrum (DKFZ, German Cancer Research Center), Heidelberg, Germany

[2] Institute of Imaging Science, Vanderbilt University, Nashville, TN 37232, USA

[3] Department of Radiology and Radiological Sciences, Vanderbilt University, Nashville, TN 37232, USA

[4] Department of Neurological Surgery, Vanderbilt University, Nashville, TN 37232, USA

[5] Department of Biomedical Engineering, Vanderbilt University, Nashville, TN 37232, USA

[6] Department of Physics and Astronomy, Vanderbilt University, Nashville, TN 37232, USA

Running Head:
**Simple spillover corrected and $T_1$ compensated CEST-MRI**

Department of Medical Physics in Radiology,

Deutsches Krebsforschungszentrum (DKFZ),

Im Neuenheimer Feld 280, D–69120 Heidelberg, Germany

*Corresponding author:
………..
German Cancer Research Center (DKFZ)
Department of Medical Physics in Radiology
Im Neuenheimer Feld 280
D–69120 Heidelberg
Germany

**List of Abbreviations**

| | |
|---|---|
| AREX | apparent exchange dependent relaxation |
| APT | amide proton transfer |
| CEST | chemical exchange saturation transfer |
| CERT | chemical exchange rotation transfer |
| CNR | contrast-to-noise ratio |
| cw | continuous wave |
| DC | duty cycle |
| EPI | echo planar imaging |
| FA | flip angle |
| gagCEST | glycosaminoglycan CEST |
| GRE | gradient echo |
| lab | label scan |
| $M_0$ | thermal equilibrium magnetization |
| MT | magnetization transfer |
| MTR | magnetization transfer ratio |
| $MTR_{asym}$ | magnetization transfer ratio asymmetry |
| $MTR_{normref}$ | magnetization transfer ratio normalized by reference scan |
| $MTR_{pcm}$ | magnetization transfer ratio of probabilistic combined model |
| $MTR_{Rex}$ | spillover corrected magnetization transfer ratio yielding $R_{ex}$ |
| $M_{z,sat}$ | z–magnetization after saturation |
| NOE | nuclear Overhauser effect |
| ref | reference scan |
| $R_{eff}$ | effective water relaxation in the rotating frame |
| $R_{ex}$ | exchange dependent relaxation in the rotating frame |
| rf | radiofrequency |
| ST | saturation transfer |
| TSE | turbo spin echo |
| WASSR | water saturation shift referencing |
| Z | normalized z–magnetization after saturation ($M_0/M_{z,sat}$) |
| $Z_{lab}$ | Z–value of label scan |
| $Z_{ref}$ | Z–value of reference scan |

# Abstract

Endogenous chemical exchange saturation transfer (CEST) effects are always diluted by competing effects such as direct water proton saturation (spillover) and macromolecular magnetization transfer (MT). This leads to unwanted $T_2$ and MT signal contributions that lessen the CEST signal specificity to the underlying biochemical exchange processes. A spillover correction is of special interest for clinical static field strengths and protons resonating near the water peak. This is the case for all endogenous CEST agents, such as amide proton transfer, –OH-CEST of glycosaminoglycans, glucose or *myo*–inositol, and amine exchange of creatine or glutamate. All CEST effects appear also to be scaled by the $T_1$ relaxation time of water, as they are mediated by the water pool. This forms the motivation for simple and novel metrics that correct the CEST signal.

Based on eigenspace theory we propose novel magnetization transfer ratio ($MTR_{Rex}$), based on the inverse Z-spectrum, which eliminates spillover and macromolecular MT effects. This metric can be simply related to $R_{ex}$, the exchange–dependent relaxation rate in the rotating frame, and $k_a$, the inherent exchange rate. Furthermore, it can be scaled by the duty–cycle, allowing for simple translation to clinical protocols. For verification, the amine proton exchange of creatine in solutions with different agar concentration was studied experimentally at clinical field strength of 3 T, where spillover effects are tremendous. We demonstrate that spillover can be properly corrected and also quantitative evaluation of pH and creatine concentration is possible. This proves that $MTR_{Rex}$ is a quantitative and biophysically specific CEST–MRI metric. Applied to acute stroke induced in rat brain, the corrected CEST signal shows significantly higher contrast between stroke area and normal tissue as well as less $B_1$ dependency compared to conventional approaches.

# Introduction

CEST exploits chemical exchange of labile protons either in metabolites or contrast agents to transfer labeled magnetization to the water pool (1–3). The CEST signal is obtained by water signal acquisition after selective rf irradiation at the resonance frequency of an exchanging proton pool. Together with a reference scan the water signal decrease due to saturation transfer can be determined. This leads to increased sensitivity mediated by the accumulation of the labeled state in the water pool. As the labeling of the exchanging protons can be done selectively by radio frequency (rf) irradiation at the specific chemical shift, CEST yields biochemical information of living tissue. Several CEST-MRI approaches were reported that enable monitoring cellular metabolites *in vivo*: amide–proton–CEST (4), creatine–CEST (5,6), glutamate–CEST (7), glycosaminoglycan–CEST (8–10), glucose–CEST (11) and also many paramagnetic exogenous agents (3). Some exchange processes are distinctly pH–sensitive and allow pH–weighted MRI (4,12–14). This makes CEST imaging interesting for characterizing of ischemic lesions as they occur in stroke where a drop of the amide proton transfer peak is reported (15–17).

However, the rf irradiation used for labeling also excites nearby resonances. Especially for clinical static field strengths and endogenous amide, amine, and hydroxyl protons, CEST pools resonate close to the water peak with the consequence that direct saturation of the water protons surpasses the CEST effect. The impact of direct water saturation on the CEST pool resonance is called "spillover". Moreover, magnetization transfer (MT) effects owing to broad macromolecular resonances are apparent, even far away from the water peak and overlay the CEST effect. Both spillover and MT effects grow with increasing rf irradiation amplitude $B_1$ (18,19). Likewise, solute labeling (and hence the possible maximum CEST effect) also increases with $B_1$ (20). Thus, CEST sequences are often optimized by variation of $B_1$ to yield maximum contrast (7,10,11), but the signal at optimal $B_1$ is highly sensitive to spillover (21) and correction is especially required in this optimal case. Finally, as mediated by the water pool, the water $T_1$ scales the strength of the whole effect.

To assess the physiological relevance of any MR imaging contrast, artifacts and sources of non-specific contrast must be identified, explained and eliminated as far as possible. This problem is an active area of research in the CEST community and several approaches were suggested to correct for the described effect: simple asymmetry analysis (4), Lorentzian line fits (22), or Lorentzian differences (23,24), or more sophisticated isolation approaches such as double frequency irradiation (25,26) and chemical exchange rotation transfer (27,28). However, even CEST signals isolated from concomitant effects, such as $T_2$ or MT, can still be diluted by them (19,20). Thus, isolation often isolates an already diluted effect by spillover and MT does generally not imply spillover and MT correction: these are different issues.

Therefore, we propose a new evaluation method which is simply applicable to Z–spectrum data and able to both isolate and correct the effects on the CEST signal from spillover, MT and $T_1$. The approach uses the inverse metric of the Z–spectrum (1/Z) to obtain spillover– and MT–corrected CEST–MRI data. Our approach is based on the equivalence of spinlock (SL) and CEST experiments (29). By employing a solution of Santyr *et al.* (30) proposed for pulsed SL, we extend this equivalence to pulsed CEST which is required in applications to clinical MR scanners. As a proof–of–principle we present data from creatine–agar model solutions, but expect that this approach is generally applicable to all types of CEST experiments driven to steady–state. We also demonstrate that the inverse metric 1/Z is useful not only for corrections, but also for quantitative CEST–MRI. Finally, we apply the correction to amide proton transfer (APT) imaging in acute stroke, where a pure exchange–weighted contrast might help in the characterization of lesions.

# Theory

### *The Z-spectrum and useful magnetization transfer ratios for cw-CEST*

We first compile results of cw theory (31,32) for two exchanging pools, the abundant pool a (water pool) with thermal magnetization $M_{0a}$ and the rare pool b (CEST pool) with thermal magnetization $M_{0b}$. For most metabolite/water systems an asymmetric population given by the proton fraction $f_b = M_{0b}/M_{0a} < 1\%$ can be assumed. Both pools undergo longitudinal and transverse relaxation with rates $R_{1a}$, $R_{2a}$, $R_{2b}$. The longitudinal relaxation of pool b, $R_{1b}$, is assumed to be small compared to the exchange rate and will be neglected in the following (32). The pools are coupled by the exchange rate $k_b$ and the back exchange rate $k_a = f_b \cdot k_b$. The rf irradiation amplitude ($\omega_1 = \gamma B_1$) and offset frequency from water ($\Delta\omega$) define the off-resonant saturation which leads to an effective field vector $\omega_{eff} = (\omega_1, 0, \Delta\omega)$ tilted by the angle $\theta = \tan^{-1}(\omega_1/\Delta\omega)$ off the z axis. (18, 22). The steady–state magnetization after saturation, $M_{sat}$, normalized by the thermal magnetization $M_{0a}$, was named Z–value by Woessner *et al.* (33). The Z–spectrum or Z(Δω) is given by (32)

$$Z(\Delta\omega) = \frac{M_{sat}(\Delta\omega)}{M_{0a}} = \cos^2\theta \frac{R_{1a}}{R_{1\rho}} \tag{1}$$

where $R_{1\rho}$ is the longitudinal relaxation rate of the water pool in the rotating frame

$$R_{1\rho} = R_{eff} + R_{ex} \tag{2}$$

$R_{ex}$ is the exchange–dependent relaxation in the rotating frame; $R_{eff}$ corresponds to $R_{1\rho}$ of the water pool when there is no exchange and reads

$$R_{eff} = R_{1a}\cos^2\theta + R_{2a}\sin^2\theta \tag{3}$$

This rate arises from the relaxation due to direct water saturation and is the origin of the spillover effect. If also a symmetric MT is apparent, $R_{eff}$ can be extended (32,34) by an exchange relaxation for MT to $R_{eff} = R_{1a}\cos^2\theta + R_{2a}\cos^2\theta + R_{ex}^{MT}$. ~~As we remove $R_{eff}$ in the following, the exact knowledge of $R_{2a}$ or $R_{ex}^{MT}$ is not important, so long as $R_{eff}$ stays symmetric in $\Delta\omega$.~~ As we remove $R_{eff}$ in the following, the exact knowledge of $R_{2a}$ or $R_{ex}^{MT}$ is not important, as long as we have a reference value with the same $R_{eff}$. In agar phantoms, $R_{eff}$ stays symmetric in Δω and the opposite frequency can be used. *In vivo*, the baseline can be estimated by a three-point method (eq. (21)) .

The exchange weighting in CEST–MRI is induced by the rate $R_{ex}$. Hence an appropriate evaluation method must provide direct access to $R_{ex}$. $R_{ex}$ can be approximated by (32)

$$R_{ex}(\Delta\omega_b) = k_b f_b \cdot \alpha \frac{\Gamma^2/4}{\Gamma^2/4 + \Delta\omega_b^2} \tag{4}$$

where $\Delta\omega_b$ is the frequency offset with respect to the CEST pool b. The labeling efficiency can be approximated by (32)

$$\alpha = \frac{\omega_1^2}{\omega_1^2 + k_b(k_b + R_{2b})} \tag{5}$$

$R_{ex}$ is a Lorentzian function of $\Delta\omega_b$ with its maximum at $\Delta\omega_b = 0$ and linewidth

$$\Gamma = 2\sqrt{\frac{k_b + R_{2b}}{k_b}\omega_1^2 + (k_b + R_{2b})^2} \tag{6}$$

A useful quantity is the magnetization transfer ratio asymmetry ($MTR_{asym}$), which attempts to isolate the contributions of CEST effects to the Z–spectra by using a reference scan without CEST effects, which can be the scan at opposite frequency or the fit of direct water saturation (23,24).

For agar phantoms, we use the opposite frequency signal as a reference. To abbreviate the following relations we define the label scan around the resonance of pool b as $Z_{lab} = Z(+\Delta\omega)$ and the reference scan at the opposite frequency with respect to water as $Z_{ref} = Z(-\Delta\omega)$. For $Z_{ref} = Z(-\Delta\omega)$, the effective relaxation $R_{eff}(-\Delta\omega)$ is unchanged, i.e. $R_{eff}(-\Delta\omega) = R_{eff}(\Delta\omega)$. Hence, $R_{ex}$ is only important for the labeling scan and $R_{ex}$ can be neglected for the opposite frequency. Thus the opposite frequency can be used as a reference scan $Z_{ref} = Z(-\Delta\omega)$. (This reasoning assumes (1) that $R_{eff}$ is symmetric and that no MT asymmetry or additional exchanging pools at $-\Delta\omega$ are present; and (2) that the width $\Gamma$ of $R_{ex}(\Delta\omega_b)$ is smaller than the chemical shift ($\Gamma < \delta_b$) of the corresponding pool.

For continuous wave (cw) steady–state it was shown previously (32) that there are different MTR normalizations possible. The most common is the substraction of Z–values of label and reference scan giving the asymmetry of the Z–spectrum:

$$MTR_{asym} = Z_{ref} - Z_{lab} = \cos^2\theta \frac{R_{ex}R_{1a}}{R_{eff}(R_{eff} + R_{ex})} \quad (7)$$

A reported spillover–corrected evaluation proposed by Liu *et al.* (35,36) normalizes with the reference value

$$MTR_{normref} = \frac{Z_{ref} - Z_{lab}}{Z_{ref}} = \frac{R_{ex}}{R_{eff} + R_{ex}} \quad (8)$$

A **p**robabilistic **c**ombined **m**odel (pcm) for Z–spectra able to separate CEST from spillover and MT of Zaiss *et al.* (22) can also be written as a magnetization transfer ratio:

$$MTR_{pcm} = \frac{Z_{ref} - Z_{lab}}{Z_{ref} - Z_{lab} + Z_{lab}Z_{ref}} = \frac{R_{ex}}{\cos^2\theta \cdot R_{1a} + R_{ex}} \quad (9)$$

This evaluation easily leads to the ideal proton transfer rate $PTR = k_a/(k_a + R_{1a})$(1,37). However, the straightforward way to separate for $R_{ex}$ is by the use of the subtraction of the inverse Z–values:

$$MTR_{Rex} = \frac{1}{Z_{lab}} - \frac{1}{Z_{ref}} = \frac{R_{ex}}{\cos^2\theta \cdot R_{1a}} \quad (10)$$

While $MTR_{asym}$ contains a quadratic term of $R_{eff}$ and $MTR_{normref}$ still a linear term of $R_{eff}$ in the denominator, note that $MTR_{pcm}$ and $MTR_{Rex}$ are free of $R_{eff}$ terms and therefore particularly free of $R_{2a}$ and symmetric MT contributions.

## *The Z–spectrum and useful magnetization transfer ratios for pulsed–CEST*

The assumption of $R_{1\rho}$ decay during the pulse (pulse duration $t_p$) and $R_{1a}$ recovery during rf off (interpulse delay time $t_d$) leads to the following formula for the steady–state in a pulsed SL experiment (30):

$$Z_{pulsed}^{ss}(\Delta\omega) = \frac{(1-e^{-R_{1a}\cdot t_d}) - \frac{R_{1a}\cdot\cos\theta}{R_{1\rho}(\Delta\omega)}(1-e^{R_{1\rho}(\Delta\omega)\cdot t_p})}{e^{R_{1\rho}(\Delta\omega)\cdot t_p} - e^{-R_{1a}\cdot t_d}} \tag{11}$$

This is the result of Santyr *et al.* (30) in our notation. Assuming small arguments of the exponential functions we employ exp(x) ≅ 1+x and obtain – with duty–cycle DC = $t_p/(t_p+t_d)$ – an expression for the normalized steady–state magnetization $Z_{SS}$ which is similar to the result for the cw case

$$Z_{pulsed}^{ss}(\Delta\omega) \approx \frac{R_{1a}(1-DC+\cos\theta\cdot DC)}{R_{1\rho}^{pulsed}(\Delta\omega)} \tag{12}$$

where

$$R_{1\rho}^{pulsed}(\Delta\omega) = R_{1\rho}(\Delta\omega)\cdot DC + R_{1a}\cdot(1-DC) \tag{13}$$

Assuming cos θ = 1 and employing eqs. (12) and (13) for the magnetization transfer rates $MTR_{pcm}$ (eq. (9)) and $MTR_{Rex}$ (eq. (10)) leads to the pulsed MTRs:

$$MTR_{pcm} = \frac{R_{ex}\cdot DC}{R_{1a} + R_{ex}\cdot DC} \tag{14}$$

$$MTR_{Rex} = \frac{R_{ex}\cdot DC}{R_{1a}} \tag{15}$$

Therefore, the pulsed MTRs are only altered by the duty–cycle as a prefactor of $R_{ex}$ against the cw case (cmp. Eqs. (9) and (10)). Thus, this spillover and MT correction is likewise applicable to pulsed CEST.

### *Quantitative Parameter Determination*

Equation 10 (with expected parameter dependencies as described in eq. (15)) provides a robust measure for qualitative contrast that avoids the effects from direct water saturation and exchange from symmetric macromolecular MT effects and is hence superior to the current standard $MTR_{asym}$. However, ideally we would like a metric that quantitatively reports on a fundamental sample parameter. To extend this spillover correction method to a quantification method we define the apparent exchange–dependent relaxation AREX

$$AREX = MTR_{Rex}\frac{R_{1a}}{DC} = R_{ex} \tag{16}$$

AREX should yield $R_{ex}$ which is given by equation ( 4) and the labeling efficiency α (eq. ( 5)). We can assume the maximum labeling efficiency (α ≈ 1) as long as the conditions $k_b << \omega_1$ and $R_2 << \omega_1$ hold (eq. ( 5)). In this full–saturation limit, when applying irradiation at the b–pool resonance, $R_{ex} = k_b \cdot f_b = k_a$ (eqs. (7) and (12)) and hence,

$$AREX = k_a \tag{17}$$

Instead of modeling a pulse train by an average $B_1$ power we now obtain a DC weighting of an average $R_{ex}$ during the pulse. The implicit assumption that the average $R_{1\rho}$ during the pulse equals the cw relaxation rate is discussed below.

In summary, two conditions must be valid for AREX = $k_a$: Firstly, the full–saturation limit $\omega_1 >> k_b$, and secondly, $R_2$ and the peak width (eq. (6)) must be smaller than the chemical shift difference to water $(\Gamma < \delta_b)$. Considering amide protons of proteins at $B_0$ = 3T, the CEST pool parameters are $k_b \approx 30\ s^{-1}$ and $\delta_b$ = 3.5 ppm (=447 ·2π $s^{-1}$) (1,38). As a model for amide exchange we employ amine protons of creatine at lower pH (6.2–6.6) to obtain a comparable exchange rate ($k_b \approx 30-60\ s^{-1}$) (19). The smaller chemical shift $\delta_b$ = 1.9 ppm (=242·2π $s^{-1}$) challenges our method for spillover correction and is therefore a crucial test. For 1 μT < $B_1$ < 5 μT both conditions are well fulfilled for creatine amine and amide protons:

$$k_b = 50\ s^{-1} << \omega_1 = \frac{B_1}{\mu T}\cdot 42.6\cdot 2\pi\ s^{-1} \cong \frac{\Gamma}{2} < \frac{\delta_b}{2} = 121\cdot 2\pi\ s^{-1}.$$

In general, if peaks are broad and other pools c are involved the reference scan is also contaminated by non–zero exchange–dependent relaxation terms $R_{ex.ref} = R_{ex}(-\Delta\omega)$ in addition to the on–resonant term $R_{ex.lab} = R_{ex}(\Delta\omega)$. Thus, AREX yields

$$AREX = R^b_{ex,lab} - R^b_{ex,ref} + R^c_{ex,lab} - R^c_{ex,ref} \tag{18}$$

instead of $R_{ex,lab}$. Although affected by the pool c and therefore being not selective anymore, the resulting $MTR_{Rex}$ is still free from symmetric spillover effects and hence an apparent $R_{ex}$ (AREX) which is still a pure exchange–dependent parameter.

## pH mapping

Under assumption of full saturation, AREX is given by $k_b f_b$. Therefore, AREX of APT can be used to calculate the exchange rate by the formula previously derived by Sun *et al.* (39):

$$pH(\textit{in vivo}) = 6.4 + \log_{10}\left(\frac{AREX}{f_b \cdot 5.57}\right) \tag{19}$$

For APT $f_b$ = 1:867 was reported (39). For amine exchange of creatine the k(pH) dependency was given by Goerke *et al.* (40). It can be rearranged yielding the absolute pH employing an AREX map:

$$pH(19°C) = \log_{10}\left(\frac{AREX}{f_b \cdot 1.4615}\right) \tag{20}$$

A flow chart of how the theory and evaluation methods are applied to the raw data is given in Figure **1**.

### *Methods*

The proposed spillover correction based on the inverse metric was tested in CEST experiments with creatine model solutions at $B_0$ = 3 T as well as *in vivo* rat measurements at 9.4T.

### Phantoms

Eleven phantoms containing phosphorus–based sodium–pottasium buffer at different pH values were measured. Their marks are shown in Figure **2** and their properties are listed in Table 1. Creatine–monohydrate (Sigma–Aldrich, Steinheim, Germany) of 55.5 mM concentration was added to each 50–ml tube. Two tubes had different creatine concentrations #F1: (2/3) × 55.5 mM and #F2: (1/3) × 55.5 mM.

To vary the conditions for spillover and MT, 0.2% to 1% agar was added to a 55.5–mM creatine solution at pH = 6.38 (#A1 – #A5, "solidified phantoms"). The pH value did not change during the heating process and addition of agar. Phantom #0 contained no agar as well as #PH1–PH3 which are controls with different pH of 6.2, 6.3, and 6.6, respectively.
Phantom parameters were estimated for a two–pool model by a full numerical fit of Z–spectra obtained for different rf amplitudes $B_1$ (numerical QUESP (41)).

## *In vitro* MRI experiments

Phantom imaging was performed on a 3–T whole–body MR scanner (Magnetom TIM–TRIO; Siemens, Erlangen, Germany). Z–spectra were obtained after saturation by a train of 80 Gaussian–shaped pulses with duration $t_p$ = 100 ms of each pulse and interpulse delay $t_d$ = 100 ms (DC = 50%) at $B_1$ = flip angle/($\gamma t_p$) = 0.2–2 µT followed by single–shot TSE imaging (FOV (220 mm)$^2$, matrix 192×192, in-plane resolution 1.1×1.1×4 mm$^3$). Z–spectra were $B_0$–corrected employing a WASSR map (42). After $B_0$ correction the MTRs according to equations (7), ( 8), (9), and (10) were calculated pixel–by–pixel and by ROI evaluation employing the opposite frequency as reference scan $Z_{ref}$. $MTR_{Rex}$ was compared to analytical $R_{ex}$ and $k_a$ values determined by the numerical fit. Fitting of Z–spectra was performed by stepwise matrix solution (33) of the 2–pool Bloch–McConnell equations.

$T_1$–weighted MR images were acquired by a saturation recovery GRE (TE = 4 ms, TR = 8 ms, 4 shots, 4 averages, FOV (220 mm)$^2$, matrix 256×256, in–plane resolution 0.9×0.9×4 mm$^3$, flip angle = 8°). Altogether 21 contrasts at different recovery times between 50 ms and 5 s were fitted to obtain $T_1$ maps.

$T_2$–weighted MR images were acquired by a spinecho sequence with 32 echo delays (TE = 11 ms to 352 ms, TR = 6 s, FOV (220 mm)$^2$, matrix 192×192, in–plane resolution 1.1×1.1×4 mm$^3$, flip angle = 180°). A pixel-by-pixel logarithmic fit was applied to obtain $T_2$ maps.

## Animal preparation

All animal–related procedures were approved by the Institutional Animal Care and Use Committee at the Vanderbilt University. The middle cerebral artery occlusion model (MCAO) was adapted on Spontaneously Hypertensive male rats (Charles Rivers Laboratory) weighing between 275 and 300 g as previously described (43). Specifically, rats were anesthetized with isoflurane (3% for induction and 2% during surgery) via a vaporizer with $O_2$. A midline neck incision was made and the common, external and internal carotid arteries were identified on the right side and isolated from the surrounding structures. The proximal branches of the external carotid artery were ligated and an arteriotomy was made in the external carotid artery. A 0.37–mm diameter silicon–coated 4–0 nylon suture (Doccol Corporation, Redlands, CA) was introduced into the vessel and routed into the internal carotid artery. The suture was pushed into the internal carotid artery until a mild resistance was felt and the MCA was occluded (at a length of 18–20 mm) and the suture was left there. Body temperature was maintained with a heating pad during surgery. The wound was then closed and buprenorphine was administered for post–operative pain management.

## *In vivo* MRI experiments

Animal imaging was performed 48 hours after surgery on a 9.4–T horizontal MRI scanner (Varian, Palo Alto, CA, USA). Bite bar and head bar were used to secure the animal during imaging to reduce respiration–induced motion artifact. The rectal temperature was kept at 37°C using a warming–air feedback system. A single–shot echo planar imaging (EPI) was used for the acquisition and a triple–reference imaging scheme (44) was used to reduce EPI artifacts. Measurement parameters were: matrix size = 64, echo time = 28 ms. Pulse train parameters were $t_p$ = 12.5 ms, $B_1$ = 0.84 µT, DC = 50%, flip angle = 180°, n = 200.

### Evaluation

For the phantom preparation creatine was employed as CEST agent and agar was used to solidify several tubes, changing the properties of the water pool. The opposite frequency is a reasonable choice for the reference scan ($Z_{ref} = Z(-3.5ppm)$) for the phantom study. However, in *in vivo* Z–spectra the opposite side is contaminated by MT and peaks owing to nuclear Overhauser effects (NOE). Therefore, for in-vivo APT signal, the three point method proposed by Jin at al. (45) was employed, in a pixel-by-pixel basis. Using the same $Z_{ref}$,

$$Z^*_{ref} = \frac{Z(3.0\,ppm) + Z(4.2\,ppm)}{2}; Z_{lab} = Z(3.5\,ppm) \tag{21}$$

we define the spillover– and MT–corrected parameter $MTR_{Rex}$

$$MTR^*_{Rex} = 1/Z_{lab} - 1/Z^*_{ref} \tag{22}$$

and, according to eq. (16), also the $T_1$–relaxation–compensated parameter AREX*. To enhance sensitivity for ROI evaluations, the MTR values of 3.4 ppm, 3.5 ppm and 3.6 ppm were averaged by using a $Z_{ref}$ from a linear interpolation of the points at 3 ppm and 4.2 ppm.

## Results

The results section is divided in three parts. First, we present the outcome of Z–spectroscopy of the solidified phantoms and how the different metrics allow for spillover correction. Secondly, we show the quantitative metric AREX and how exchange rate and pH mapping can be obtained after compensation of effects of $T_1$–relaxation. Thirdly, the metrics are applied to *in vivo* data of stroke in rat brain.

### Spillover correction

Figure 3a shows distinct effects in the Z–spectrum upon addition of 1% agar (phantom #A5) in experiments at $B_0 = 3$ T. The corresponding asymmetry is strongly diluted compared to the creatine solution without agar (phantom #0) ($MTR_{asym}$, Figure 3b). Even in the case of full saturation of the CEST pool ($B_1 > 1$ μT) we still observe a strong dependence of $MTR_{asym}$ on $B_1$ in the solidified phantom. $MTR_{normref}$ (Figure 3c) provides an enhancement of the signal of the solidified phantom, but displays an underestimation of the CEST effect by about 30 % and a strong dependence on the amplitude $B_1$ of the saturating field. Both $MTR_{pcm}$ and $MTR_{Rex}$ give very similar values for solidified and non–solidified phantoms. This can be seen not only in the case of on–resonant irradiation on pool b, but also the shape of the CEST peak is coherent with the peak of the control measurement in the absence of agar. This proves the validity of our spillover correction for arbitrary frequency offsets. In the case of full saturation the dependence of $MTR_{Rex}$ and $MTR_{pcm}$ on $B_1$ is less than 15%, producing a small overestimation of the effect. Near the center of the water proton resonance at 0 ppm all MTRs show considerable deviations which are even larger for the inverse $MTR_{pcm}$ and $MTR_{Rex}$. Errors increase tremendously which is discussed in detail below (Figure 10).

Images of $MTR_{asym}$ and $MTR_{Rex}$ at 1.83 ppm display the same relation: Spillover is uncritical for small $B_1$, whereas for stronger $B_1$ spillover dilutes $MTR_{asym}$ significantly. In contrast, $MTR_{Rex}$ yields an homogeneous contrast up to $B_1 = 1.4$ μT independent of the agar concentration (Figure 4). Figure **4** demonstrates the importance of spillover correction since $MTR_{asym}$ causes misinterpretation of diluted signals as changes in pH or concentration. At $B_1 = 2$ μT the agar tubes #A2 and #A3 show a very similar contrast as the tube #F2 in presence of one third of the creatine concentration. For exceedingly high $B_1$, the contrast–to–noise ratio is insufficient and $MTR_{Rex}$ cannot completely reconstruct the ideal signal from the residual signal. $MTR_{Rex}$ also enhances the signals from the tubes without agar: different pH and different concentration can therefore even better be distinguished after correction.

Figure **5** displays MTR as a function of $B_1$. Using $MTR_{Rex}$ or $MTR_{pcm}$ the plateau of the full–saturation limit is reached, while $MTR_{asym}$ and $MTR_{normref}$ show the known decrease of the CEST effect due to spillover dilution induced by MT, $T_2$ relaxation, and $B_1$. Thus in the full–saturation limit a spillover correction is also a first–order $B_1$ correction.

## Quantification

The numerical Bloch–McConnell fit of ROI–averaged Z–spectra for different $B_1$ yields the characterization of the phantom parameters listed in Table 1. The values for exchange rates $k_b$ agree well with WEX data measured by Goerke *et al.* (40). The relative concentrations $f_b$ are in good agreement with the prepared creatine concentration if the number of exchanging protons per molecule is 4. $R_{2b}$ values are quite constant with a value of ca. 50 $s^{-1}$, which is comparable to the exchange rate.

$MTR_{Rex}$ and $MTR_{pcm}$ were calculated using $B_1$, DC, $k_b$, $f_b$, $R_{1a}$, $R_{2b}$, and the theoretical rate $R_{ex}$ (eq. ( 4)). Actually, the pulsed approach of Santyr et al. (30) is known to deviate especially for slow exchange rates (46). Roelloeffs et al. (46) showed that a biexponetial decay during the break has to be modeled to properly extend the model of Santyr. For spillover and MT correction we think this is not important as the deviation is still only exchange dependent, for quantification this may have an influence. Nevertheless, the comparison of $MTR_{Rex}$ calculated from fit results with $MTR_{Rex}$ obtained from data (Figure **5**) shows that corrected curves can still be interpreted by the analytical solution for $MTR_{Rex}$ (eq. (10)(15)) based on Santyrs model. We think that because we used 100ms Gaussian pulses - which end with a low power - the equilibrium between pool a and pool b is not too much changed directly after the pulse, making the biexponantial decay less important. However, this should be studied in detail and may limit our approach for pulsed CEST of slow exchanging systems saturated with different pulse shapes. However, with our pulse parameters the step from spillover–corrected $MTR_{Rex}$ to a reliable quantification of the back exchange rate $k_a$ is straightforward by employing eq. (17) (Figure **6**). AREX is therefore proportional to the concentration $f_b$ and the exchange rate $k_b$. It varies between phantoms #F1, #F2 and #0 and also between phantoms #0, #PH1, #PH2, and #PH3. AREX yields homogeneous contrast in phantoms #0,#A1–A5. However, a small overestimation of $k_a$ in the agar phantoms compared to the control (#0) is observed.

Using the exchange rate of creatine protons, $k_b$(pH = 6.38, T = 19°C) = 35 $s^{-1}$, measured in WEX experiments (40) , a map of the relative proton fraction $f_b$ can be obtained which is valid for the given pH and is in agreement with fit results. Together with the prepared creatine concentration (55.5 mM) this approach yields the number of labile protons per creatine molecule, $N = f_b \times [2H_2 0]/[Cr]$. For pH ≤ 6.4, N is most probably 4, in conformity with the zwitterionic structure of creatine (Figure **2**b) and the $pK_a$ value of the creatine amine groups $pK_a$ = 6.6 at T = 37°C (5). For the phantom at pH=6.6 the proton number might be smaller. Assuming 4 exchanging amine protons for creatine the value of $f_b$ can be derived for any creatine concentration. Together with AREX = $k_a$ we obtain a $k_b$–map. Finally, using the dependence of the creatine amine exchange rate $k_b$ on pH found by Goerke *et al.* (40), a map of absolute pH values can be calculated. The resulting data is in good agreement with pH values prepared in the phantoms (Figure **6**). This proves that spillover, MT, and $T_1$–relaxation compensation worked well for the creatine–agar phantoms.

## Application *in vivo*

Having demonstrated the validity of the introduced corrections, the formalism can be applied to *in vivo* data. In a stroke lesion of rat brain we expect a drop of APT due to pH drop, which is clearly visible in Figure **7**. The Z–spectrum at 3.5 ppm is contaminated by direct saturation and MT effects (Figure 7d,g), therefore the baseline estimation of Jin *et al.* (45) was employed as a reference. After correction of spillover by $MTR_{Rex}$ the contrast to noise ratio (CNR) between normal and lesion tissue increases from CNR=1.17 to CNR=1.44 (values correspond to ROIs in Figure 8). Also the $T_1$–map (Figure **7**e) shows a difference between lesion and healthy tissue (Figure **7**b). This can be corrected by the AREX evaluation showing an even higher CNR=1.62 between normal and pathologic tissue (Figure **7**c). Please note that delay time and rotation transfer effects were taken into account by using DC=1 to calculate AREX. Finally, employing equation (19) and the reported proton fraction $f_b$ = 0.115% a absolute pH map can be calculated from AREX. It shows pH values between 7 and 7.2 in normal tissue and a drop to around 6.5 within the lesion (Figure **7**f).

A further check of the spillover correction is possible by investigating the behavior for increasing $B_1$ which was shown for phantoms in Figure **4**. A similar signature in the MTRs as a function of $B_1$ was observed after spillover correction of ROI averaged data (Figure 8b, c): For low $B_1$ APT*, $MTR_{Rex}$*, and AREX* show an increase with $B_1$. After reaching a maximum at 1.6 μT signals drop again. However, the decrease for the spillover–corrected methods is less significant and a kind of plateau is reached. Again, the APT contrast after spillover correction is shown to be less $B_1$–dependent. Important to note is the increase of contrast between tissue in the lesion and normal tissue. For AREX the contrast difference is much larger than the standard deviation. Therefore, AREX leads to a more pure, but also larger contrast.

## Discussion

In this study we showed that a magnetization transfer ratio employing the inverse metric of the Z–spectrum enables removal of spillover and MT effects from CEST signals.

As depicted in Figure 1 only simple mathematical operations are needed to get a spillover–corrected signal from raw Z–spectra data. Previous studies on spillover by Sun *et al.* (19,20,47) treated the spillover effect by introducing a spillover coefficient σ of the ideal magnetization transfer rate, i.e. $MTR_{real} = (1-\sigma)\cdot\alpha\ MTR_{ideal}$ (α is the labeling efficiency, eq. (7)). In contrast to this approach, we observed that spillover dilution can be better explained by the inverse addition of contributing effects. Spillover dilution of a CEST effect induced by "parallel" saturation of water resembles the "dilution" of a resistor $R_b$ by a parallel circuit to another resistor $R_a$. If the diluted resistance $R_{a+b}$ and the resistance of $R_a$ are known, one obtains $1/R_b = (1/R_{a+b}) - (1/R_a)$.

The reason why superposition and not inverse superposition of effects in the Z–spectrum were also successful in other treatments results originates from the approximation $Z = 1/(1+x) \approx (1-x)$ valid for $R_{eff} \approx R_{1a}$ and small $x \approx R_{ex}/R_{1a}$. This is also the principal reason why superpositions of Lorentzians can be fitted to steady–state–pulsed CEST spectra. According to our results a superposition of reciprocal Lorentzians should be more suitable. The observation that 1/Z yields basically an $R_{1\rho}$ spectrum (eq. (1)) further supports the importance of the inverse Z–spectrum. $R_{1\rho}$, known from spinlock experiments, has, as a relaxation rate, the property of being a superposition of the apparent exchange–dependent relaxation effects (eq. (2) (48)).

Some degree of dilution was identified as spillover effect by Sun *et al.* (19,20,47). In our approach this contribution is regarded as a loss in labeling efficiency. The latter can be defined more generally as $\alpha = R_{ex}/k_a$ yielding (29):

$$\alpha = \underbrace{\frac{\frac{(\omega_b-\omega_a)^2}{\omega_1^2+\Delta\omega^2}k_b + R_{2b}}{k_b+R_{2b}}}_{*}\frac{\omega_1^2}{\omega_1^2+k_b(k_b+R_{2b})} \tag{23}$$

A comparison of eq. (23) with $\alpha$ of eq. ( 5), which is similar to the $\alpha$ given in of Ref. (37), reveals an additional factor (*) in eq. (23), which decreases with increasing $B_1$. This factor is maximal at $\Delta\omega = 0$ and can be interpreted as on–resonance effects induced by the exchange. Those effects are employed in on–resonant spinlock experiments. The loss of labeling is attributed to an interference of off–resonant and on–resonant features of $R_{ex}$. Hence, labeling efficiency is an useful parameter which was extended by the eigenspace approach (29), whereas a spillover coefficient is not appropriate to the inverse metric.

Other than the spillover correction employed by Sun *et al.* (19) and the "isolation" of $R_{ex}$ from $R_{1\rho}$ ($R_{ex} = R_{1\rho}-R_{eff}$) proposed by Jin *et al.* (14), which both use additional $T_2$– and $B_1$–mapping, our approach employs only the intrinsic metric to correct spillover. This is advantageous because it reduces scanning time and post–processing efforts. The intrinsic structure was employed in a similar manner in the fit model of Ref. (22) which is the origin of $MTR_{pcm}$ (eq. (5)).

It is important to note that isolation and correction of effects are different operations. Whereas the former approaches like $MTR_{asym}$, isolate signals from specific effects, these isolated effects can still be diluted (Figure 3). Therefore, removing the information about parallel effects must be considered carefully, since other contributions may become invisible, but can still be effective as dilutions.

The approach of Liu *et al.* (35) (eq.( 8)) afforded a partial spillover correction which could be explained by the reduction of the quadratic term of $R_{eff}$ in $MTR_{asym}$ to a linear one. Their normalization is considered to be appropriate for glutamate CEST imaging (7,49) and gagCEST (50). In particular, these two applications can also benefit from the improvement of the $MTR_{Rex}$ evaluation.

Next we discuss our quantitative approach. To the best of our knowledge, there are two different approaches to model the pulsed–CEST case. On the one hand, using cw theory with an equivalent cw power (12,51) verified to be valid for slow exchange rates by Tee *et al.* (52). On the other hand, the approach of Santyr *et al.* (30) for spinlock which should also be valid for CEST by relying on the equivalence of spinlock and CEST (29). Santyrs spinlock solution takes into account different relaxation during and between the pulses. However, Santyr *et al.* assume solely monoexponential decay in the interpulse delay which cannot explain the modulations as a function of the flip angle observed by chemical exchange rotation transfer (CERT) (28) or the dependency on delay time (53). However, for long pulses, as employed in this study, our results suggest that Santyrs approach is also valid for pulsed CEST. We observe no perfect match of Santyrs theory and the corrected data, what we attribute to the invalid assumption of monoexponential dynamics in the interpluse delay and the assumption of full saturation ($R_{ex} = k_a$) during the Gaussian pulse. The $R_{ex}$ obtained by AREX is therefore an effective parameter which incorporates the dependence on pulse shape as well as processes occurring between the pulses.

The inverse metric is valid only for pulsed CEST/SL if $R_{1\rho}{\cdot}t_p << 1$ (assumption of eq. (12)) which is not the case for large $R_2$ and $\theta$ or $t_p$. This could explain why the agar phantoms show slightly different $MTR_{Rex}$ compared to the solutions without agar (Figure 4). In principle, this limits the inverse approach to applications with pulses that are much shorter or much longer than $1/R_{1\rho}(\Delta\omega)$ The latter corresponds to the cw case. In practice, Z values are directly tunable by $B_1$ and can be set to values that are not smaller than 0.5; then $R_{1\rho} \sim 2\,R_{1a}$ and the condition $R_{1\rho}{\cdot}t_p << 1$ is easier to fulfill.

Pulsed CEST including MT was also studied with similar phantom parameters by Desmond *et al.* (18) who could interpret their data with numerical Bloch-McConnell simulations. By addition of agar $T_2$ is changed strongly. However, a limitation of our study is that MT was only shown to be corrected up to 1% agar, which corresponds to a fraction of about $f_b = 0.3\%$ (54). In contrast, fractions up to $f_b = 18\%$ are possible in cartilage (55). For cw simulations showed that the inverse superposition is appropriate up to $f_b = 5\%$ (29), but then the assumption of the simple superposition $R_{1\rho} = R_{eff} + R_{ex,CEST} + R_{ex,MT}$ might be invalid.

The intermediate exchange regime was not explicitly considered in this study. In this case the spillover correction of $MTR_{Rex}$ is promising, but has still to be proven. Although demonstrated so far for amine and amide exchange, we expect that our normalization will work for DIACEST and PARACEST in the slow– and intermediate exchange regime, and for the generation of qualitative contrast and quantitative parameter fittings.

**Application in the case of non steady-state and inversion pulses**

For in-vivo protocols the saturation times are commonly kept shorter, in the range of $1 \cdot T_1$-$2 \cdot T_1$, to save scanning time or avoid dominant spillover effects (6,7). Also, the more efficient inversion pulses are commonly used (23,24,51,53). Additional measurements in non-steady-state with only 3s irradiation (~$1xT_1$) were performed (Figure 9abe), as well as saturation with a pulse train of 180°-pulses (Figure 9cdf) for the phantom described in Figure 2. In both cases the homogeneity between the agar phantoms was improved by the inverse evaluation $MTR_{Rex}$ against $MTR_{asym}$. From theory it is known (32), that the inverse metric is not valid for transient state. However, near steady–state ($t_{sat}T_1$) it can still be used as an approximation. For inversion transfer or CERT there is no analytical knowledge, but our results indicate that the general Z-spectrum structure might be similar also for rotation transfer.

**Systematic and statistical errors**

Figure **10** depicts the increase of errors for $MTR_{Rex}$. If we turn to the 1/Z metric the relative errors stay similar $\Delta(1/Z)/(1/Z) = \Delta Z/Z$, but the absolute errors increase: $\Delta(1/Z) = (1/Z^2) \cdot \Delta Z$. For example, $R_1 = 1\ s^{-1}$ and a dominant direct saturation at the label frequency of $R_{eff} = 2 \cdot R_1$ leads to $Z = 0.5$. Hence, the statistical error of 1/Z is 4 times the error of Z.

However, $MTR_{asym}$ has also a quadratic term of $R_{eff}$ in the denominator, therefore $MTR_{asym} \sim 1/Z^2 \cdot PTR$. Thus, the CEST effect estimated by $MTR_{aysm}$ has a systematic error in the order of $1/Z^2$. This means, by the inverse metric we trade systematic errors against statistical errors, which can be reduced by averaging. This also indicates that $B_1$ should not exceed a certain limit to keep Z and the SNR large. For the estimation of SNR, $MTR_{asym}$ is a good indicator.

Figure **8** indicates, that a spillover correction is also a $B_1$ correction near the full–saturation limit ($\alpha \approx 1$) . Where $MTR_{asym}$ shows a strong dependence on $B_1$ and has to be corrected by postprocessing as proposed by Sun *et al.* (47) $MTR_{Rex}$ is nearly constant up to $B_1 = 2\ \mu T$. For faster exchange and partial saturation the tissue–dependent $B_1$ correction of Singh *et al.* (56) reported at $B_0 = 7$ T should be performed with spillover–corrected $MTR_{Rex}$ instead of $MTR_{normref}$. We do not recommend to apply a $B_1$ correction on spillover–diluted data, but suggest to apply $B_1$ corrections to $R_{ex}$ directly.

**Imaging of stroke**

In the rat CEST imaging study of Sun *et al.* (15) APT of normal tissue was reported to be 2.94% whereas in the lesion it dropped to 0.9%. This was stronger than the signal decrease of about a factor of 0.5 observed in our animal study. Also in pH mapping we only see a drop of 0.5 pH units while a pH–decrease of approx. 1 was reported in ref. (15). In this and other studies (17,39) also the correlation with lesions detected by diffusion and perfusion imaging was investigated. In our experiments resolution was too low to resolve significant substructures within the lesion. In contrast to other studies we avoided contaminations of asymmetry analysis by employing a baseline estimation (45). We think that this is beneficial, especially because significant NOE effects and shifted macromolecular MT effect are apparent in the brain parenchyma. Note that this method can only be applied in higher fields ($B_0 \geq 3$ T). For lower $B_0$ we suggest a Lorentzian–line fit of the water resonance as appropriate reference.

# Conclusion

We propose a new spillover– and MT–correction method for evaluation of Z–spectra from CEST experiments which needs no information about $T_2$ and MT of the system and is easily applied. Validity of the proposed corrected magnetization transfer ratio ($MTR_{Rex}$) was demonstrated for an *in vitro* system yielding high spillover i.e. creatine in agar gels at clinical field strengths. $MTR_{Rex}$ was extended to a $T_1$-relaxation-compensated metric, called AREX for apparent exchange-dependent relaxation, which allowed quantitative evaluation of Z-spectra and could be verified by numerical fits. Validity, sensitivity and performance of the metric require sufficiently large Z-values ($Z>0.5$) and a proper reference scan. APT-CEST MRI experiments of acute stroke in rat brain at $B_0$ = 9.4 T fulfilled these requirements. The outcome of the evaluation by means of the AREX metric was a significantly higher contrast between stroke area and normal tissue compared to the contrast obtained by use of the non-inverse metric. Hence we propose application of the AREX metric for analysis of Z-spectra data of all pathologies where changes of MT, $T_2$, or $T_1$ are observed, in particular in tumors or tissue affected by stroke. With a proper reference scan AREX may help to provide a pure exchange-dependent and exchange-site-specific contrast.

**Table 1: Employed phantoms as depicted in Figure 2 with fitted values from ROI evaluations and exchange rates of creatine predicted by WEX studies (40). Fits by a two pool Bloch–McConnell simulation were performed simultaneously for 5 Z–spectra in each ROI obtained with $B_1$ = 0.2 µT, 0.4 µT, 0.6 µT, 0.8 µT, and 1.0 µT. Crosses (×) mark missing fit data when the two–pool model was insufficient.**

| Phantom /ROI | pH | [Cr] [mM] | [Ag] [%] | $k_b$ (WEX) [$s^{-1}$] | $k_b$ (CEST) [$s^{-1}$] | $f_b$ [%] | $R_{2b}$ [$s^{-1}$] | $R_{2a}$ [$s^{-1}$] | $T_{1a}$ [s] |
|---|---|---|---|---|---|---|---|---|---|
| 0 | 6.38 | 55.5 | 0 | 35.1 | 33.8 | 0.22 | 40.3 | 0.60 | 2.92 |
| A1 | 6.38 | 55.5 | 0.2 | 35.1 | x | x | x | 2.47 | 2.99 |
| A2 | 6.38 | 55.5 | 0.4 | 35.1 | x | x | x | 3.57 | 2.97 |
| A3 | 6.38 | 55.5 | 0.6 | 35.1 | x | x | x | 4.82 | 2.92 |
| A4 | 6.38 | 55.5 | 0.8 | 35.1 | x | x | x | 6.23 | 2.87 |
| A5 | 6.38 | 55.5 | 1.0 | 35.1 | x | x | x | 7.58 | 2.83 |
| F1 | 6.38 | 55.5/3 | 0 | 35.1 | 38.5 | 0.073 | 58 | 0.42 | 3.00 |
| F2 | 6.38 | 55.5·2/3 | 0 | 35.1 | 46.5 | 0.13 | 53 | 0.42 | 2.99 |
| PH1 | 6.21 | 55.5 | 0 | 23.72 | 31.6 | 0.18 | 49 | 0.44 | 2.91 |
| PH2 | 6.32 | 55.5 | 0 | 30.6 | 34.8 | 0.20 | 48 | 0.43 | 2.93 |
| PH3 | 6.61 | 55.5 | 0 | 59.6 | 63.0 | 0.21 | 54 | 0.46 | 2.93 |
| water | 7 | x | x | x | x | x | x | 2 | 3 |

Figure Captions

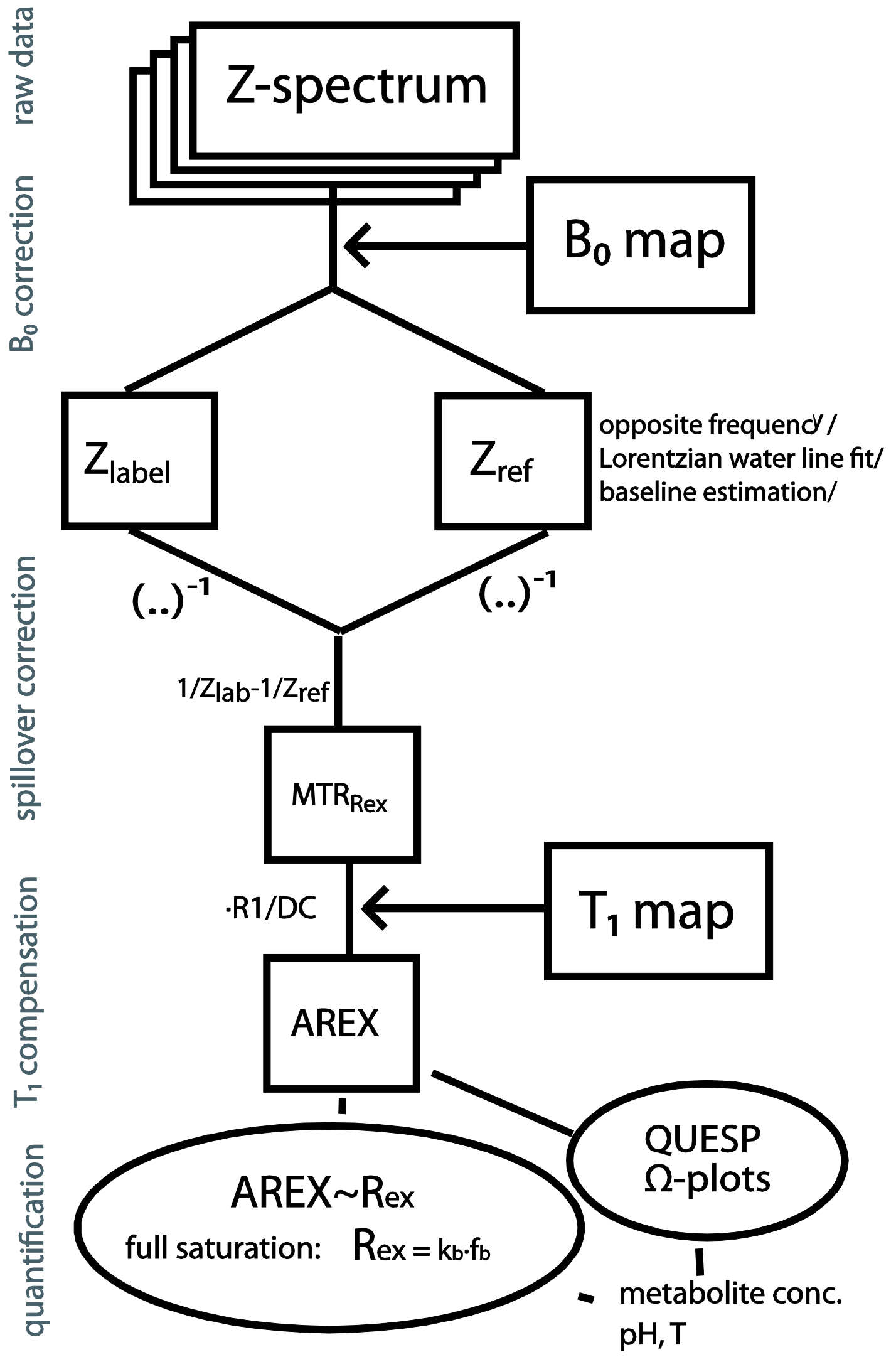

Figure 1

: Scheme of data evaluation for spillover– and $T_1$–compensated CEST. Only simple matrix operations are performed to get the AREX contrast. The step of defining a suitable reference value $Z_{ref}$ is crucial.

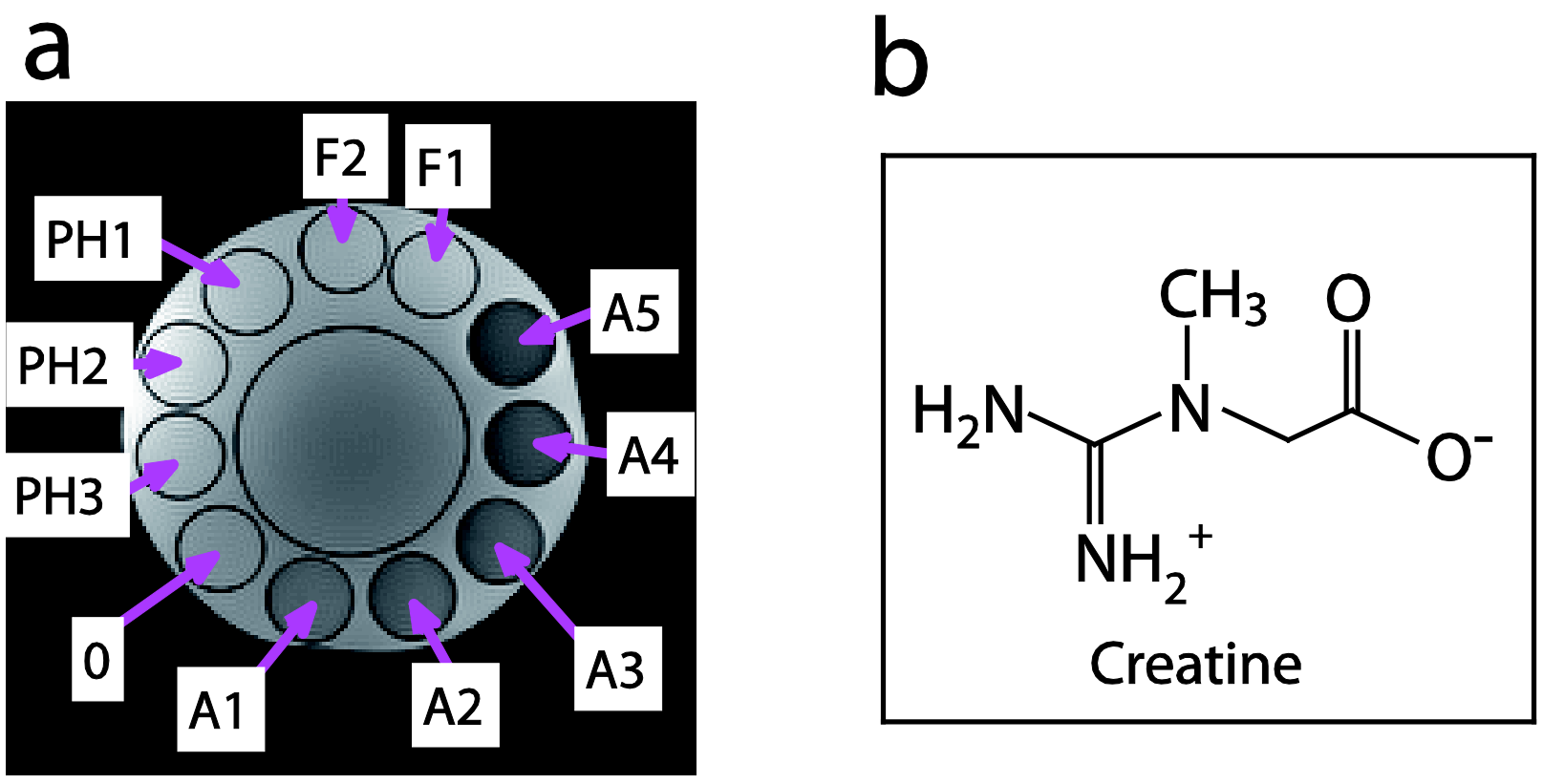

Figure 2

: a) Turbo–spinecho (TSE) image of the employed phantom (for details see Table 1). #0 is the reference solution with 55.5 mM creatine and PBS at pH = 6.38 and without agar. The #Ax phantoms differ from #0 by increasing agar concentration (0.2% – 1%). The #PHx phantoms differ from #0 by altered pH (6.2, 6.3, 6.6). The #Fx phantoms have different creatine concentrations compared to #0 (#F1:55.5 mM 1/3, #F2: 55.5 mM 2/3). (b) zwitterionic form of aqueous creatine occurring at low pH.

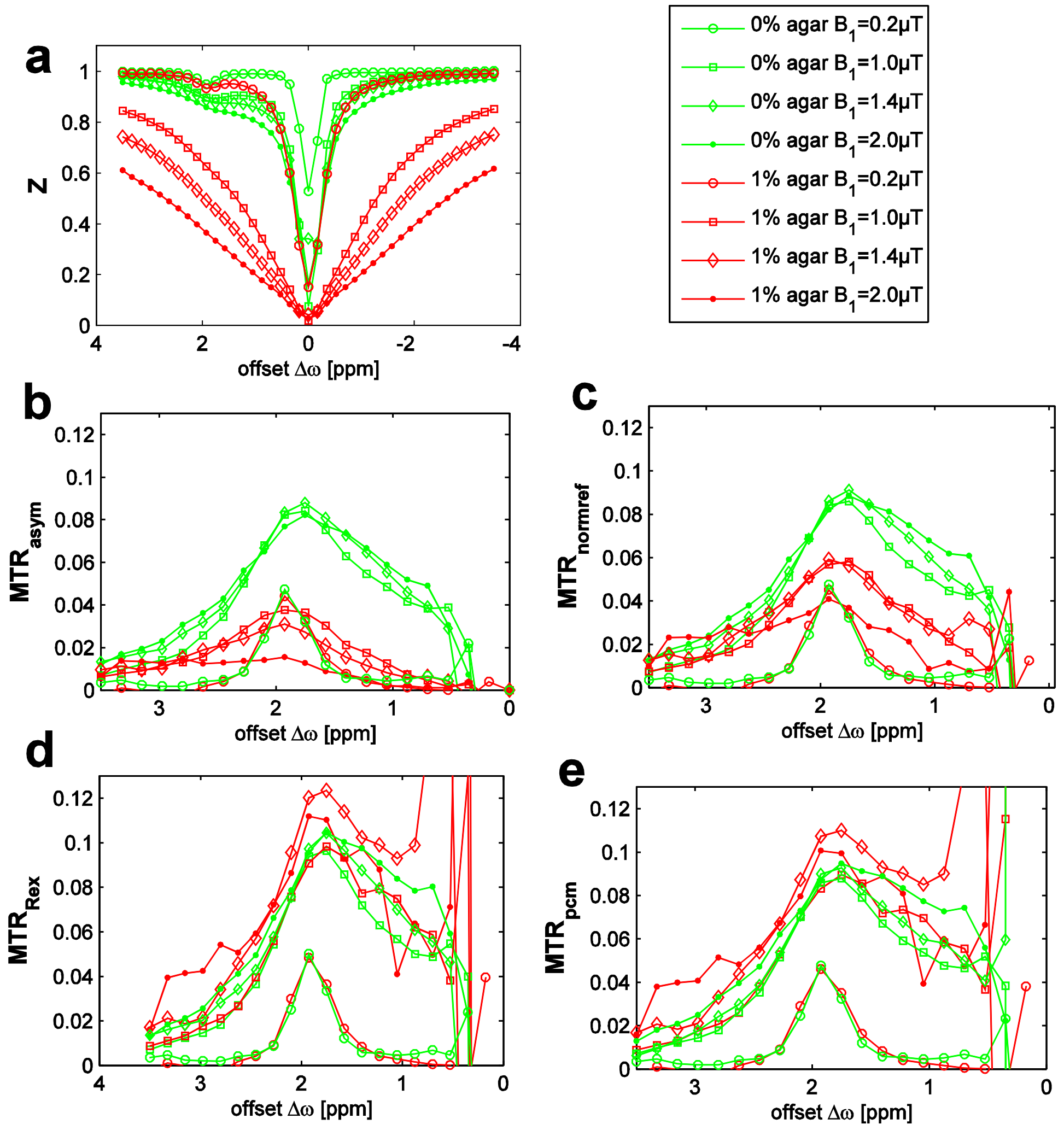

Figure 3

(a) Z–spectra obtained with different $B_1$ from creatine solutions with 1% agar (blue lines, #A5) and 0% agar (red lines, #0). The labeling increases with $B_1$, but also the direct saturation effect: for low $B_1$ of 0.2 μT (circles) the spillover effect is negligible (a) which explains why curves overlap for all metrics (b–e). (b) $MTR_{asym}$ shows the strong spillover dilution in the solidified phantom (with agar) whereas $MTR_{Rex}$ (d) and $MTR_{pcm}$ (e) are able to correct the dilution so that aqueous and solidified phantoms yield almost the same effect. The spillover correction proposed by Liu *et al.* (c) (eq. ( 8) in Ref.(35)) compensates spillover partially. Error bars are omitted for better visibility; they increase strongly for higher spillover correction as depicted in Figure 10.

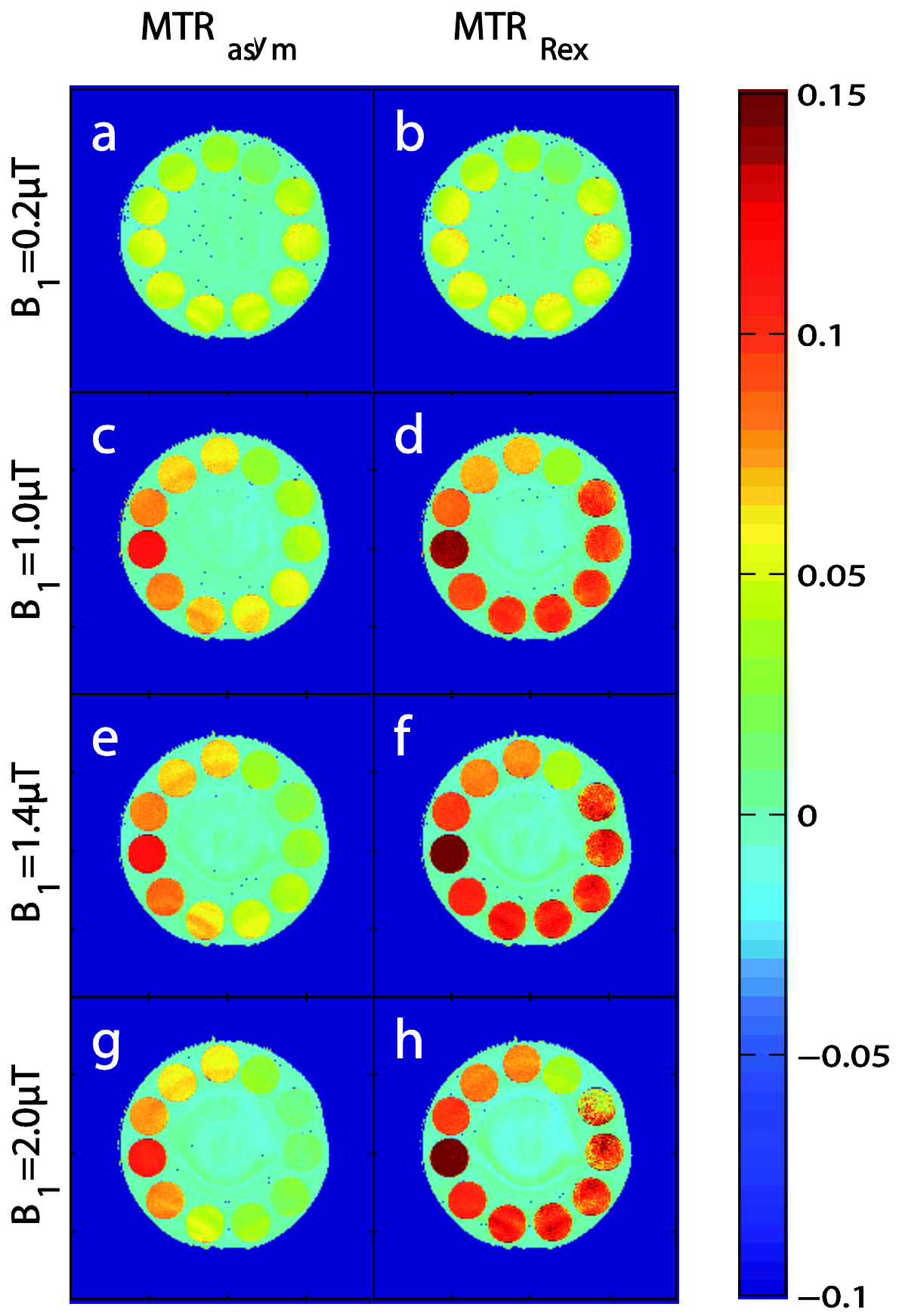

Figure 4
Comparison of the normalizations $MTR_{asym}$ (left column) and $MTR_{Rex}$ (right column) for $B_1$ = 0.2, 1.0, 1.4, and 2.0 µT. In each case the estimation of the CEST effect is higher for $MTR_{Rex}$. The phantoms with varying agar concentration (#0, #A1–#A5) show similar contrast in $MTR_{Rex}$, whereas $MTR_{asym}$ shows diluted contrast with increasing agar concentration. Differences in pH and creatine concentration are reflected in both MTRs. Therefore, $MTR_{Rex}$ has all properties of a spillover correction.

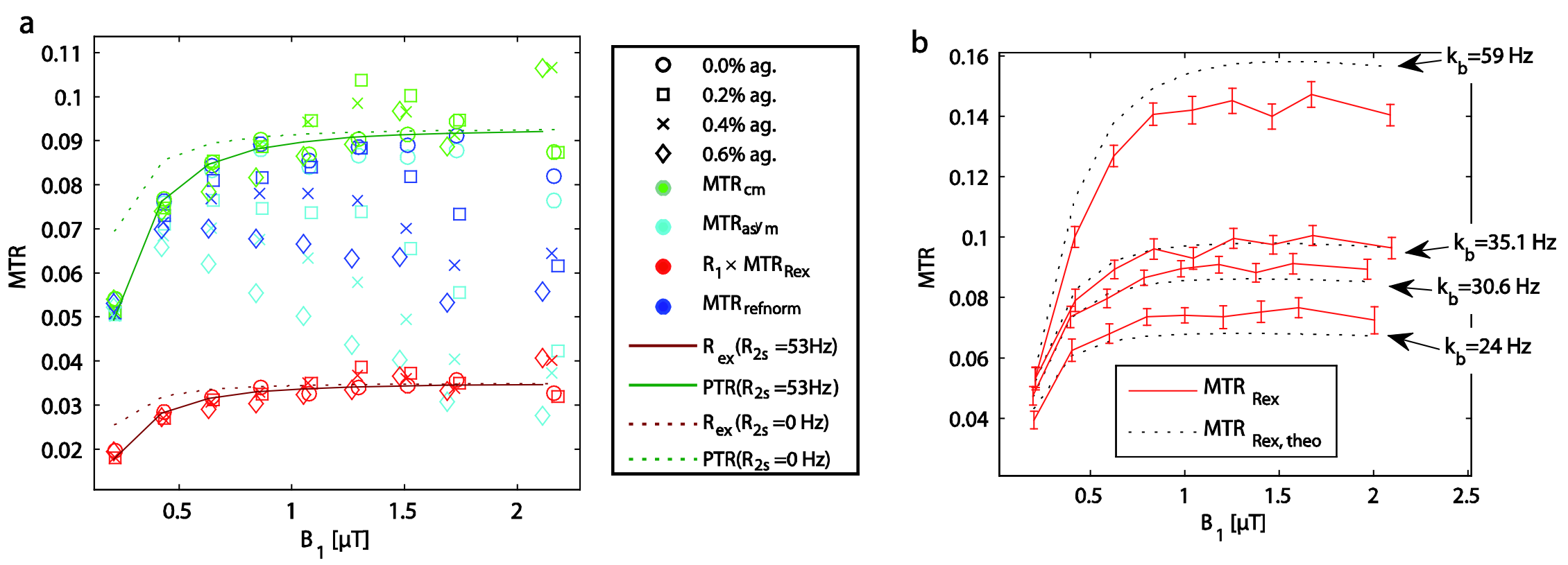

Figure 5
(a) ROI evaluation of ROIs #0 and #A1-#A3 with the proposed spillover corrections. For all agar concentrations and $B_1$ values , both $MTR_{pcm}$ (green) and $MTR_{Rex}$ (red) appear to be in a narrow band around the control without agar #0. $MTR_{asym}$ (cyan) and $MTR_{normref}$ (blue) show a much stronger decrease with increasing $B_1$ and agar. (b) $MTR_{Rex}$ from data and from theory (eq. (15)) employing parameters of the numerical fit (Table 1): The curves match roughly and the dependence on $k_b$ and $B_1$ is very similar.

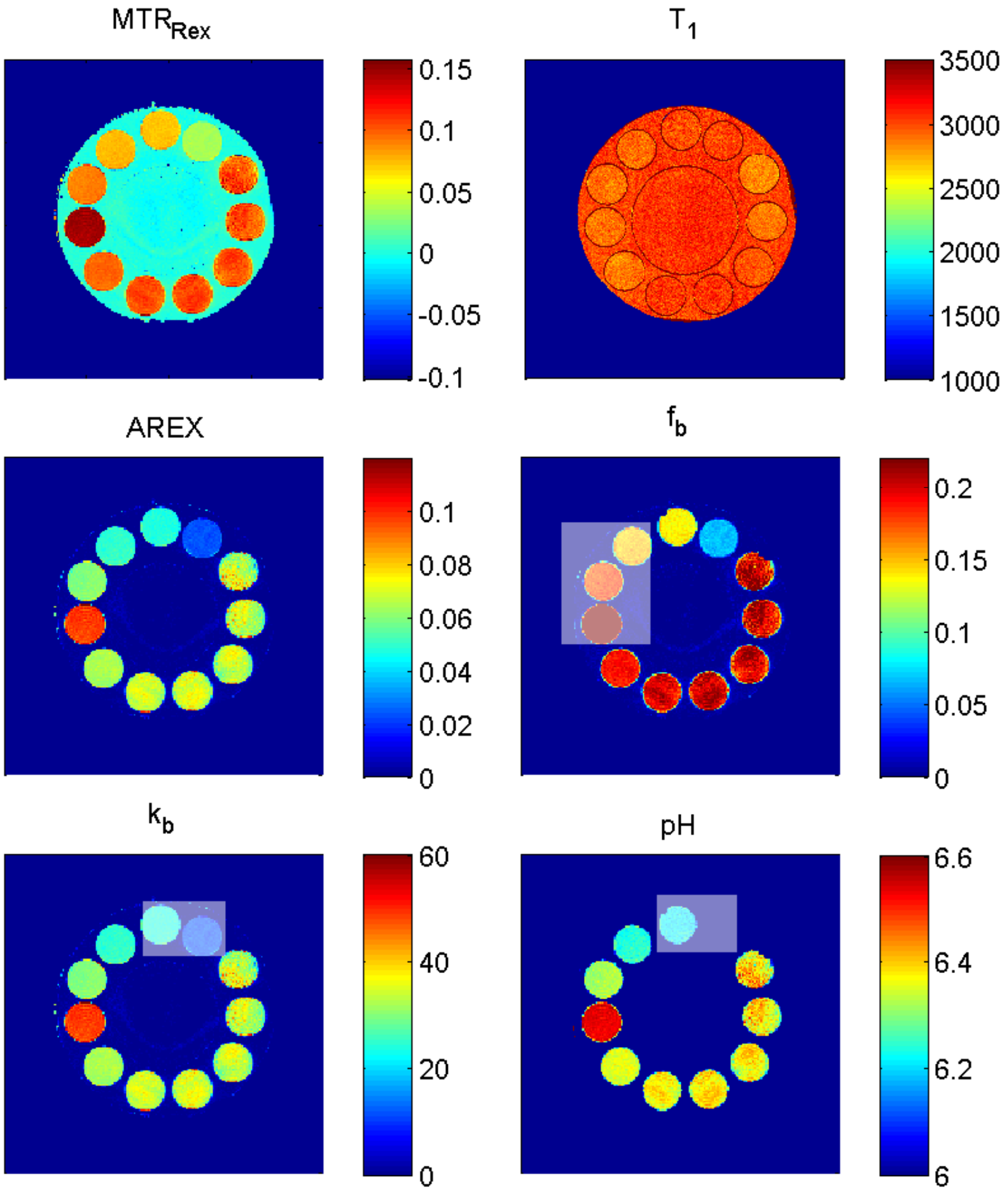

Figure 6

Quantitative pulsed CEST–MRI. (a) $MTR_{Rex}$ evaluated for $B_1 = 1$ μT. Employing the $T_1$ map (b) the spillover–corrected and $T_1$–compensated AREX map can be calculated. Under the assumption of full saturation AREX yields a $k_a$–map. (c) $f_b$–map employing the exchange rate for creatine k(pH = 6.38, T = 19°C) = 35 $s^{-1}$; it suggests that creatine has 4 exchanging protons. Using $f_b$ = 0.2% a $k_b$–map (e) can be obtained from AREX which correlates well with results from WEX measurements. (f) Therefore, a pH($k_b$)–map can be obtained using Eq. (16). Gray boxes indicate tubes were either concentration or pH was not constant.

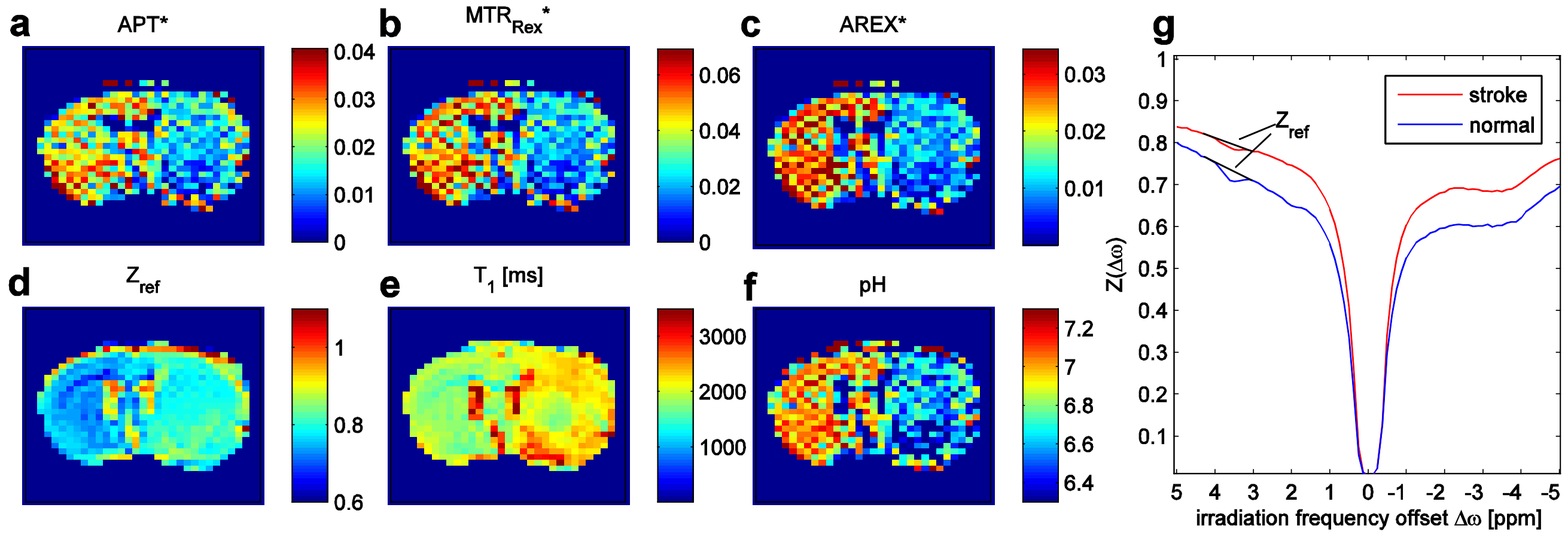

Figure 7

: Amide proton transfer (APT) contrast obtained by pulsed CEST–MRI of rat with acute stroke. APT contrast (a) is contaminated by $T_1$ (e) and spillover effects (visible in the reference image (d) ). After correction of spillover by the inverse metric $MTR_{Rex}$ the contrast between lesion and normal tissue increases (b). The $T_1$–corrected AREX evaluation yields a pure exchange–weighted contrast which shows even higher signal drop in the stroke lesion compared to normal tissue. (e) From AREX an absolute pH map can easily be obtained by eq. (19). For all CEST maps the average of Z–values at 4.2 and 3 ppm was employed as a reference (eq. 21) as illustrated by the baselines in the Z-spectra in (g). To achieve good visual comparison of the contrast each MTR map was windowed from zero to two times the average value of all non–zero pixels. CEST-EPI parameters were: matrix size = 64, echo time = 28 ms. Pulse train parameters were $t_p$ = 12.5 ms, $B_1$ = 0.84 µT, DC = 50%, flip angle = 180°, n = 200.

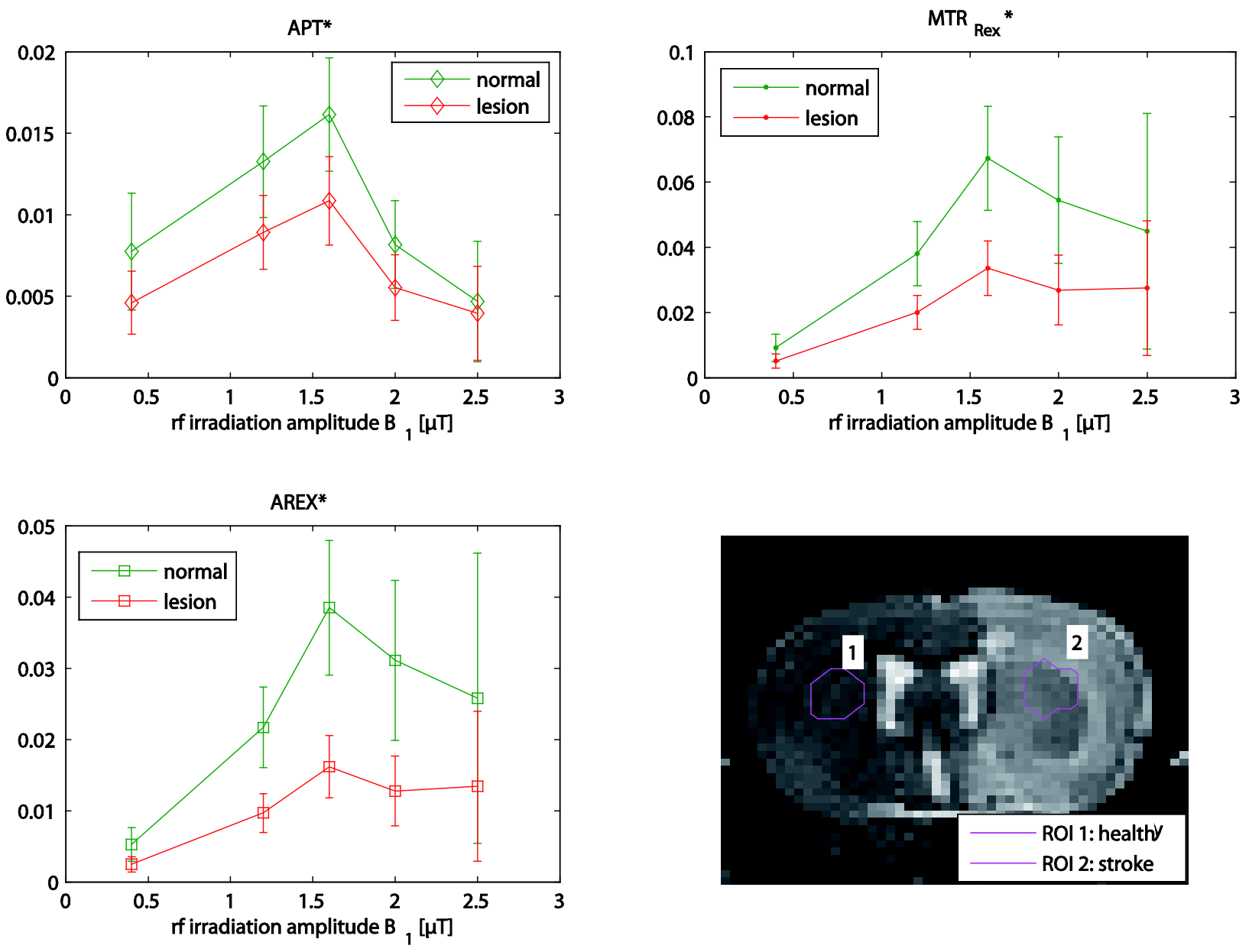

Figure 8

: ROI–evaluation of the three–point methods APT*, the spillover–corrected $MTR_{Rex}$, and spillover– and $T_1$–compensated AREX. Similar to the phantom study APT* shows a strong decrease with $B_1$ for values higher than 1.5 uT, whereas $MTR_{Rex}$ and AREX show less decrease of signal for higher $B_1$. This pattern is similar to the phantom results (compare Figure 5) and indicates validity of the spillover correction. However, the plateau of full–saturation limit is not reached, caused by probably by contaminations of the reference scan. Important to note is that the difference between signals in the stroke lesion and normal tissue are much more significant after spillover correction and $T_1$ compensation. Under assumption of equal amide concentrations in stroke and normal tissue, the signal drop reflects a change in exchange rate of about a factor of two.

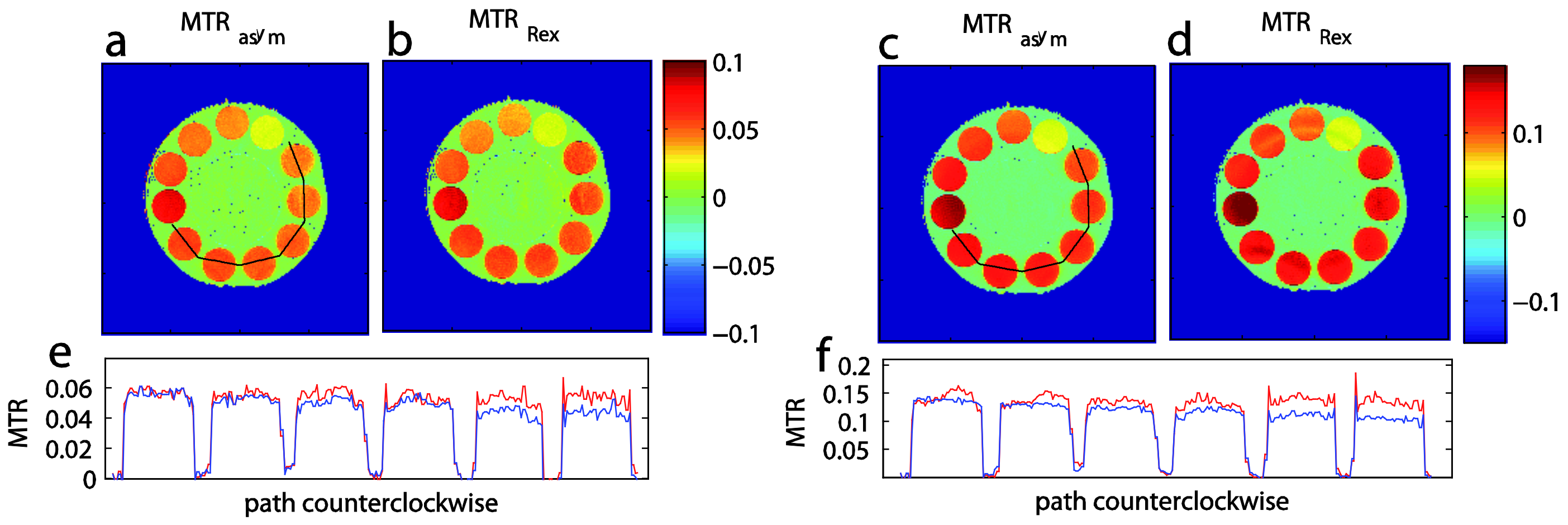

Figure 9:

$MTR_{asym}$ (a,c) and the inverse approach $MTR_{Rex}$ (b,d) for the practical relevant cases of (a,b) non-steady state saturation ( $t_{sat}$=3.1 s, $B_1$ = 0.5 µT, $t_p$ = 100 ms, DC=50%, n= 16) and (b,d) 180° pulsed saturation (180°–pulses: $B_1$ = 0.48 µT, $t_p$ = 25 ms, DC=50% n= 320 ). Profiles along the path (counterclockwise) defined in (a) show that $MTR_{Rex}$ (red line) corrects the decrease of $MTR_{asym}$ (blue line) with increasing agar concentration.

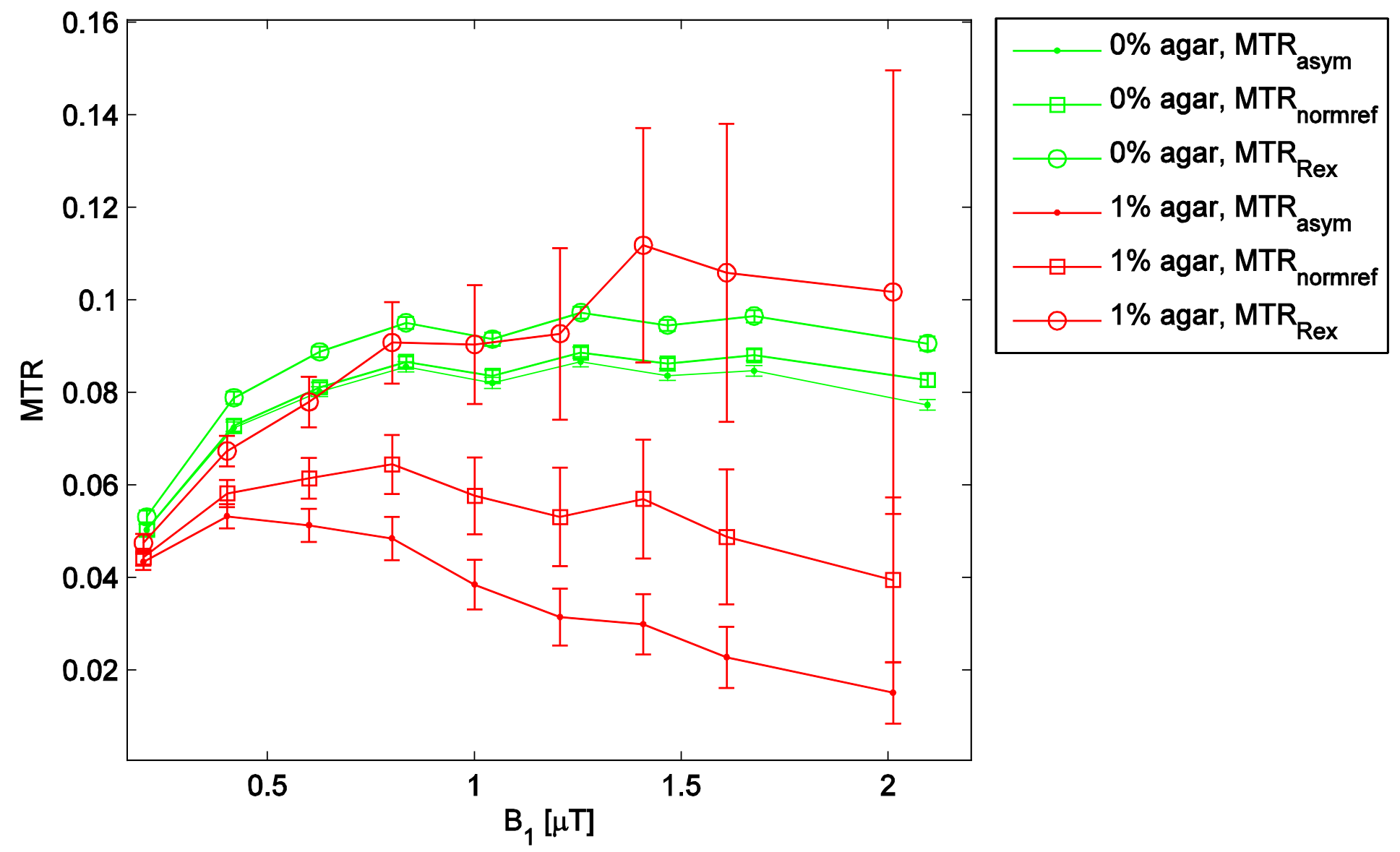

Figure 10

: Error estimation of inverse metric. Absolute errors increase with increasing spillover effect. This results from the error propagation of the inverse metric. However, relative errors do not change and therefore contrast–to–noise ratio is not affected. The systematic spillover deviation is traded with a statistical fluctuation. Original errors were scaled by a factor of one third to improve visibility.

Graphical abstract

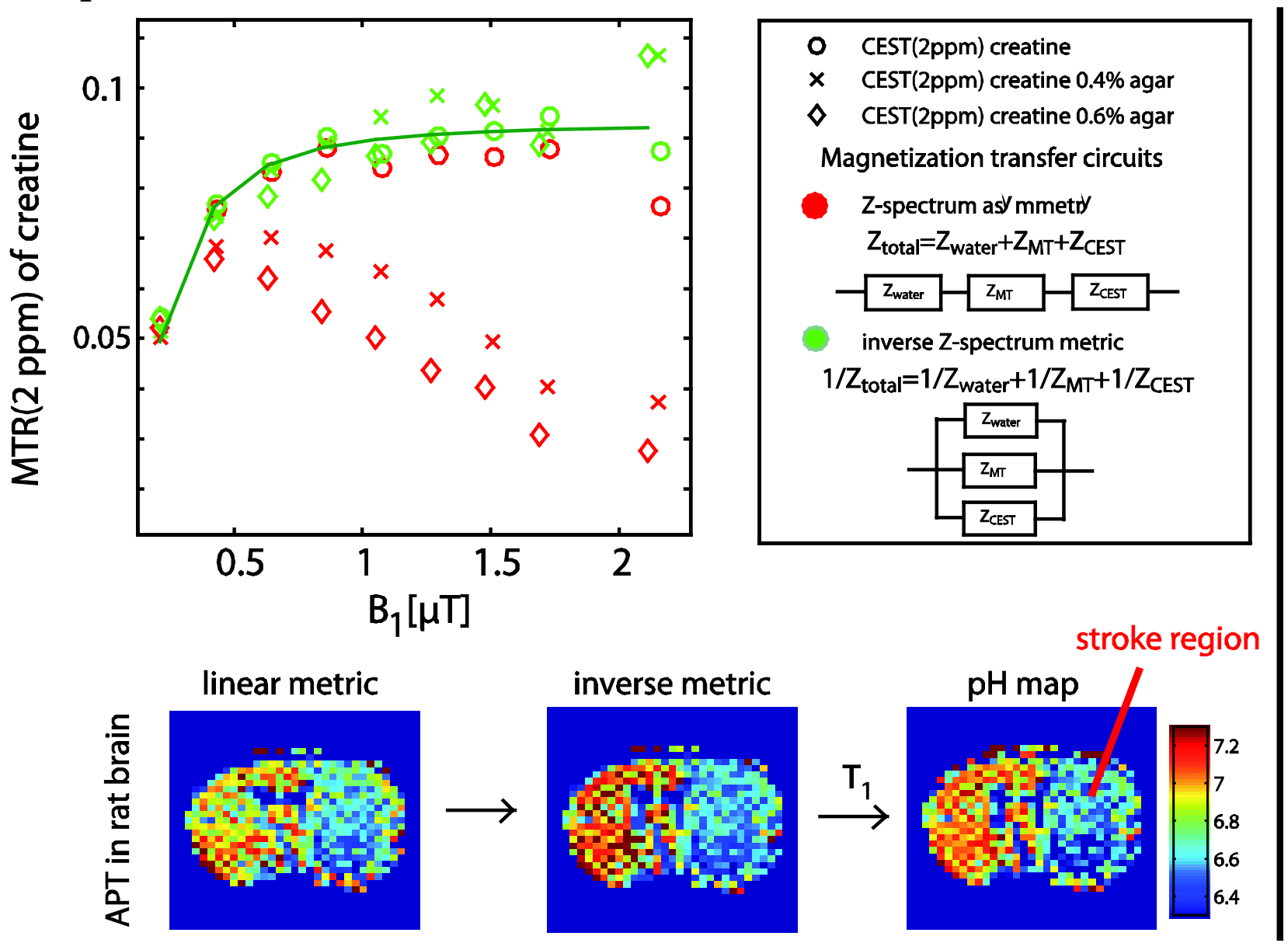

**Inverse Z-spectrum analysis for MT- and spillover-corrected and T1-compensated steady-state pulsed CEST-MRI – application to pH-weighted MRI of acute stroke**

The different effects that influence the Z–spectrum behave like resistors in a parallel circuit and not in a serial circuit. The spillover effect can thus be removed by using the inverse metric $MTR_{Rex}$. $MTR_{Rex}$ yields the exchange dependent relaxation and allows clean CEST effect evaluation which is applied to spillover- and MT- corrected and $T_1$-compensated ph -weighted amide proton transfer imaging of acute stroke.

M.Zaiss, J.Xu, S.Goerke, I.S.Khan, R.J.Singer, J.C.Gore, D.F.Gochberg, P.Bachert